\documentclass[twocolumn]{revtex4}

\usepackage{graphicx}
\usepackage{dcolumn}
\usepackage{bm}
\newcommand{\etal}{\emph{et al.}}
\newcommand{\be}{\begin{equation}}
\newcommand{\ee}{\end{equation}}
\newcommand{\bfig}{\begin{figure}}
\newcommand{\efig}{\end{figure}}
\newcommand{\incl}{\includegraphics}
\begin{document}      
\setcounter{page}{1}
\title{GEOMETRY AND THE ANOMALOUS HALL 
EFFECT IN FERROMAGNETS\footnote{Summer School on Condensed Matter Physics 2005, Princeton University }
}
\author{N. P. ONG and WEI-LI LEE$^{\dagger}$
}
\affiliation{
\mbox{Department of Physics, Princeton University, New Jersey 08544, U.S.A.}
}


\begin{abstract}
The geometric ideas underlying the Berry phase and the modern viewpoint of Karplus and Luttinger's theory of 
the anomalous Hall effect are discussed in an elementary way.  We briefly review recent 
Hall and Nernst experiments which support the dominant role of the KL velocity term in ferromagnets.
\end{abstract}

\maketitle                   
\section{Introduction}
Geometry crops up in physics in unexpected ways.  The example in this talk traces an idea
proposed 50 years ago.  Mired in controversy from the start, it simmered for a long time as an unresolved
problem, but has now re-emerged as a topic with modern appeal.  
In 1954, Karplus and Luttinger (KL)~\cite{Karplus} discovered the earliest instance of 
the Berry phase~\cite{Berry} in solids when they calculated the charge current of 
electrons moving in a Bravais lattice with broken time-reversal symmetry (a ferromagnet).  
They uncovered a quantity $\bf \Omega(k)$ that acts like a ``magnetic field'' in reciprocal 
space to produce a transverse velocity $\bf E\times \Omega$.  The velocity leads to a 
dissipationless Hall current which, according to KL, accounts for the anomalous Hall effect (AHE) 
seen in all ferromagnets~\cite{Karplus,Adams}.  Because Berry-phase physics lay far into the future, the KL theory 
encountered stiff resistance.  From the 60's to late 70's, Smit's 
rival skew-scattering theory~\cite{Smit} gained ascendancy, with apparent support from experiments 
on dilute Kondo systems, e.g. Cu$Mn$ (in hindsight, these experiments are irrelevant because 
the host lattice retains time-reversal invariance).  

In the past 4 years, interest in topological currents stimulated by 
theories of the integer and fractional Quantum Hall effects has led to intense re-examination of the KL theory from
the modern viewpoint~\cite{Niu,Nagaosa,Jungwirth,Murakami,Yao,Haldane}.  
Experimental tests~\cite{Lee1,Lee2} now provide strong evidence that the KL theory provides the 
correct explanation of the AHE in ferromagnets.  

In a Hall experiment on a ferromagnet, the observed Hall resistivity $\rho_{xy}$ is the sum
of the AHE resistivity $\rho'_{xy}$ and the usual Lorentz term $R_0H$ ($R_0$ is the ordinary Hall coefficient).  We have
\be
\rho_{xy}(T,H) = R_0(T)H + \rho'_{xy}(T,H).
\label{rhoxy}
\ee
At each $T$, the strong $H$ dependence of the AHE term simply reflects the rotation of all 
domains into alignment.  This is expressed as $\rho'_{xy}(T,H) = R_s(T)M(T,H)$ where
the anomalous Hall coefficient $R_s(T)$ is the scale factor between the magnetization $M$ and $\rho'_{xy}$.  
Our interest is on the value of $\rho'_{xy}$ strictly in the impurity scattering regime.

In this talk, we introduce the geometric ideas underlying the KL theory and review recent 
experiments.  [Readers familiar with holonomy may skip to Sec. \ref{anomalous}.]

\section{Geometry}\label{geometry}
Imagine that a beetle moves a unit vector $\bf v$ along a closed path $C$ on a curved 
surface (e.g. a latitude on the globe) with the constraint that ${\bf v}(t)$ must always lie in 
the local tangent plane, i.e. ${\bf v\cdot {e}}_3 = 0$ where ${\bf {e}}_3$ is the local 
normal vector (Fig. \ref{sphere}a).  Along the path, we impose the further restriction that 
the direction of $\bf v$ is perturbed \emph{minimally} (Fig. \ref{sphere}b).  Clearly, $\bf v$ cannot 
remain absolutely parallel to itself; the next best thing is to ensure that $\bf v$ 
never rotates about ${\bf {e}}_3$, i.e. any change $d\bf v$ must be parallel to $\pm{\bf {e}}_3$ (Fig. \ref{sphere}c), viz. 
\be
{\bf {e}}_3\times d{\bf v} = 0.
\label{dv}
\ee

\bfig[h]			
\incl[width=6.5cm]{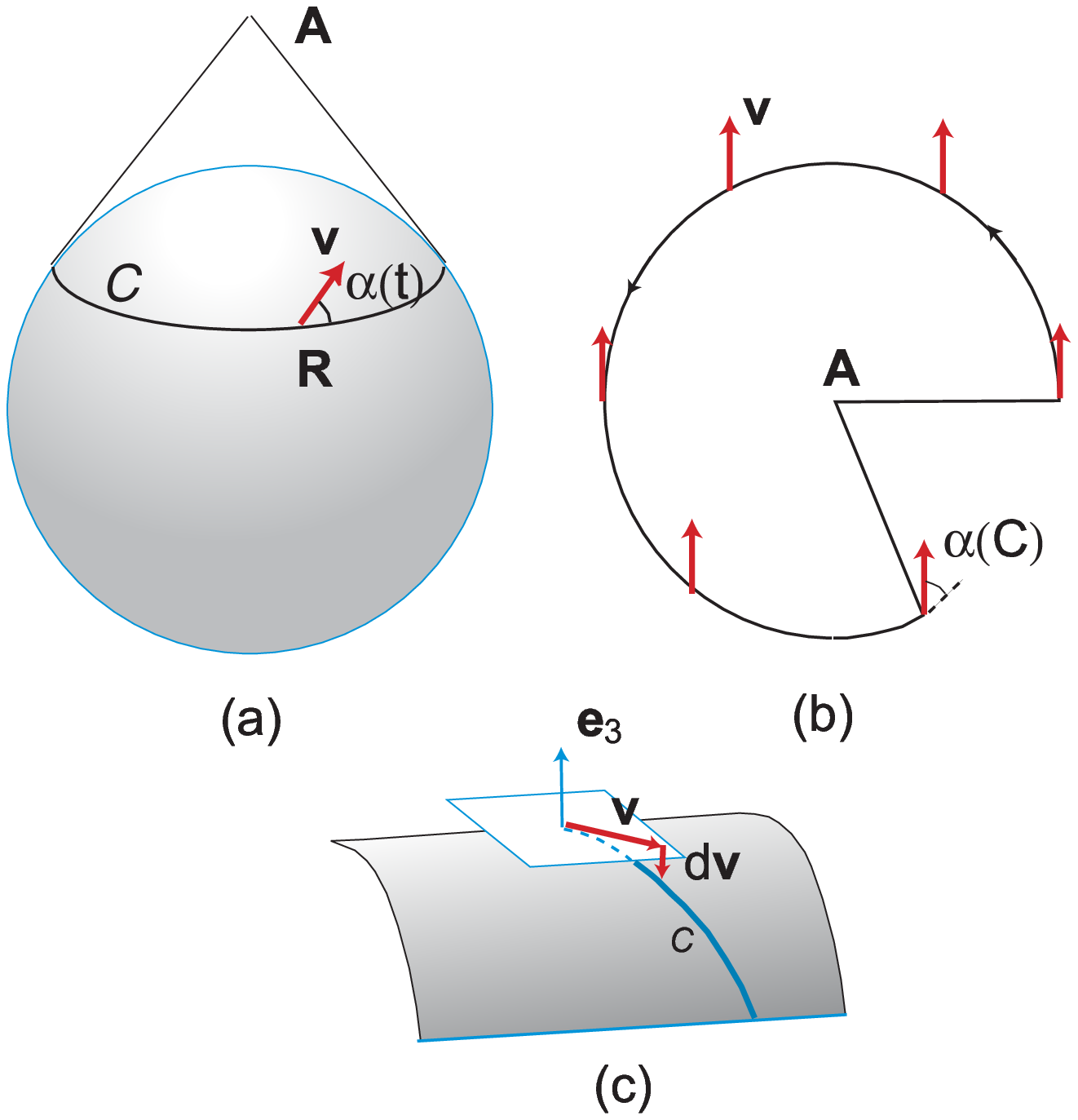}
\caption{\label{sphere}
(a) Transport of $\bf v$ along the curve $C$ on a sphere.  If $C$ is a latitude, the
tangent planes are given by the cone with apex $A$.  (b) Flattening the cone on a table
shows that $\bf v$ remains parallel to its initial direction if it is prevented from rotating
around the local normal vector ${\bf e}_3$ (parallel transport).  However, measured 
relative to the latitude, $\bf v$ makes an angle $\alpha(C)$ on returning to the 
starting point (holonomy).  (c) In general, $\bf v$ lies in the tangent plane
(white rectangle).  In parallel transport, the change $d{\bf v}$ must not have any component in 
the tangent plane, i.e. $ \pm{\bf e}_3 || d{\bf v}$. 
}
\efig

The process satisfying Eq. \ref{dv} is known as ``parallel transport'' (or Levi-Civita transport)~\cite{Stoker}.  
If $C$ is a circle on the sphere, the tangent planes are part of a cone (Fig. \ref{sphere}a).  By flattening
the cone (Fig. \ref{sphere}b) we see why the no-rotation condition is termed parallel transport.
In the ``flattened'' space, $\bf v$ is always fixed in direction.  However, on the curved surface, the beetle 
finds that $\bf v$ does not return to its initial direction when the path is completed.  Instead, it makes 
a ``geometric'' angle $\alpha (C)$ that is path dependent (holonomy, Fig. \ref{sphere}b).

To measure $\alpha$, we need a local coordinate frame [${\bf  {e}}_1(t)$, ${\bf  {e}}_2(t)$], 
where $t$ parametrizes $C$ (on the sphere, these are usually the unit vectors 
along the latitude and longitude).   At each point ${\bf R}(t)$, we expand 
\be
{\bf v}(t) = \cos\alpha(t)\;{\bf  {e}}_1(t) + \sin\alpha(t)\;{\bf  {e}}_2(t).
\label{vt}
\ee
As we move along $C$, the triad $[{\bf v,w,e}_3]$ 
(where ${\bf w} = {\bf e}_3\times {\bf v}$) rotates about the normal by $\alpha(t)$,
relative to the local frame $[{\bf e}_1,{\bf e}_2,{\bf e}_3]$.  We note that Eq. \ref{dv} 
implies 
\be
{\bf e}_3\times d{\bf w} = 0.
\label{dw}
\ee
Using Eq. \ref{vt} in Eq. \ref{dv} with the normalization 
conditions ${\bf  {e}}_1\cdot d{\bf  {e}}_1 = 0$ \emph{etc}, 
we find the elegant result
\be
d\alpha = {\bf  {e}}_1\cdot d{\bf  {e}}_2 \equiv \omega_{21}.
\label{da}
\ee
On completing the trip, the total angle $\alpha(C)$ is the integral of 
the ``connection form'' $\omega_{21}$ which encodes the overlap between
$d{\bf  {e}}_2$ and ${\bf  {e}}_1$~\cite{Stoker}.

Holonomy seems less mysterious if we regard ${\bf v}$ as the quasi-fixed direction relative to which the local frame 
rotates as we traverse $C$.  This is evident for the
flattened cone in Fig. \ref{sphere}b.  

These results may be written more compactly using complex vectors 
with the inner product ${\bf (a, b)} = {\bf a}^*\cdot{\bf b}$.
We write ${\bf \hat{n}} = [{\bf {e}}_1 + i{\bf {e}}_2]/\sqrt{2}$ 
and ${\bf\hat{\psi}} = [{\bf v} + i{\bf w}]/\sqrt{2}$.  Equation \ref{vt} becomes $\hat{\psi}(t) 
= {\bf \hat{n}}(t)\mathrm{e}^{-i\alpha(t)}$, with the angle $\alpha(t)$ now 
appearing as a phase.  The parallel transport conditions Eqs. \ref{dv}, \ref{dw} 
and \ref{da} now have the compact forms
\be
{\bf\hat{n}}^*\cdot d{\bf\hat{\psi}} = 0, \quad\quad d\alpha = -{\bf \hat{n}}^*\cdot id{\bf \hat{n}}.
\label{psi}
\ee

\section{Berry Phase}\label{berry}
In the Berry-phase problem~\cite{Berry} a parameter (the coordinate $\bf Q$ of the nucleus) is slowly
taken around a closed curve $C$, while the bound electron is assumed to remain in the same 
eigenstate $|n,{\bf Q}\rangle$ parametrized by $\bf Q$ (Fig. \ref{hilbert}a).  This assumption 
is a constraint analogous to that on $\bf v$ in Sec. \ref{geometry}.  When the circuit is
completed, the electron ket $|\psi\rangle$ does not return to its starting state $|n,{\bf Q}_{0}\rangle$.
Instead, it acquires a phase $\chi(C)$ called the Berry phase.  We now compare
this system with the previous example.  

The path $C$ traced by ${\bf Q}(t)$ lies on a surface in parameter space  $\bf Q$.
At each $\bf Q$, the eigenstate $|n,{\bf Q}(t)\rangle$ may be regarded as the local 
coordinate frame defined in Hilbert space (in place of the tangent plane), i.e. $|n,{\bf Q}(t)\rangle$
is analogous to ${\bf \hat{n}}(t)$ in Eq. \ref{psi}.  As $C$ is traversed, the ket of the 
electron $|\psi\rangle$ is parallel transported, and hence acquires 
a phase angle relative to $|n,{\bf Q}\rangle$,
viz. $|\psi\rangle = |n,{\bf Q}\rangle\mathrm{e}^{-i\chi}$ (Fig. \ref{hilbert}b).

\bfig[h]			
\incl[width=6.5cm]{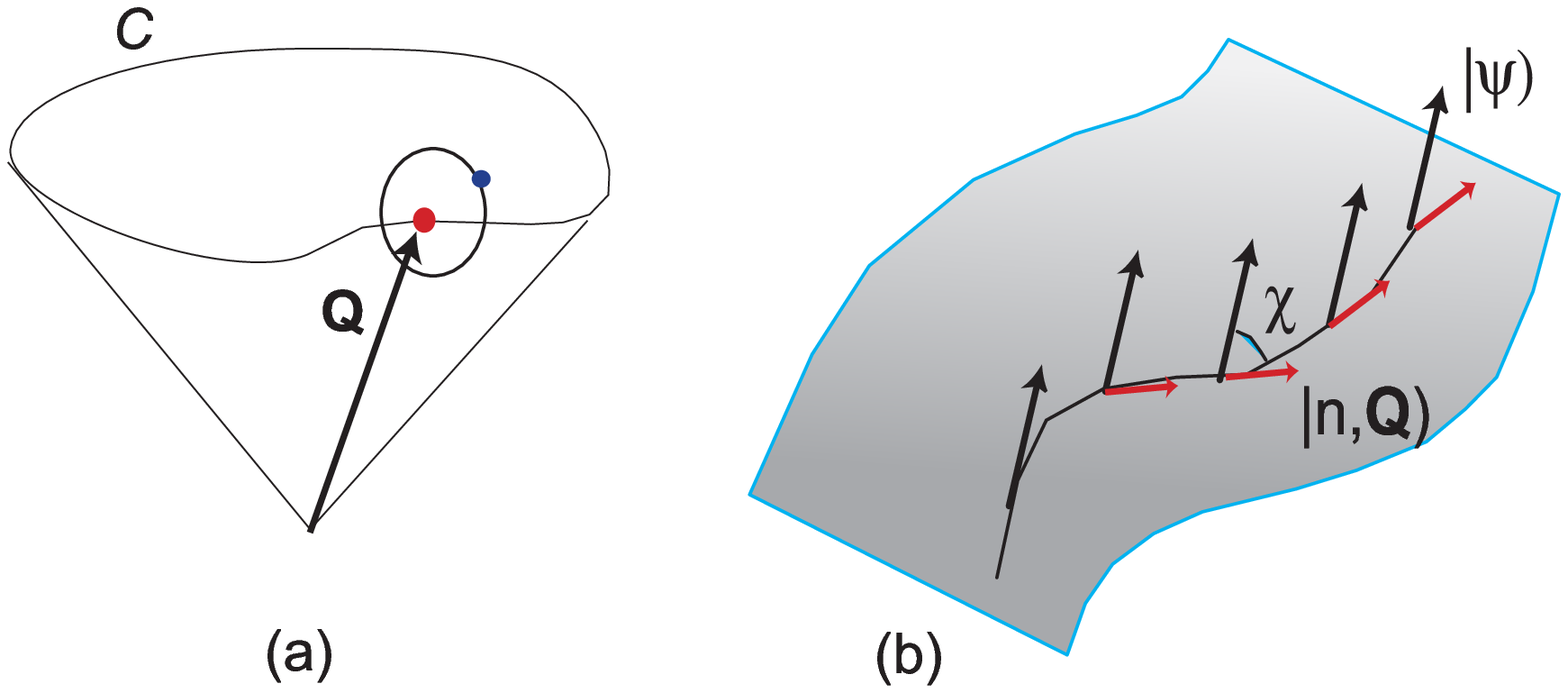}
\caption{\label{hilbert}
(a) Electron constrained to same eigenstate $|n,{\bf Q}\rangle$ as nuclear coordinate ${\bf Q}$ is 
taken around a closed path $C$. (b) Parallel transport of ket $|\psi\rangle$ in Hilbert space.
As $\bf Q$ changes, $|\psi\rangle$ acquires a phase angle relative to $|n,{\bf Q}\rangle$.
}
\efig 

Imposing the parallel-transport condition $\langle n,{\bf Q}|\delta|\psi\rangle = 0$ (Eq. \ref{psi}),
we find $\delta\chi = -\langle n,{\bf Q}|i\delta|n,{\bf Q}\rangle$.
On completing the path $C$, the total Berry phase $\chi(C)$ is the line integral
\be
\chi(C) = -\oint_C d{\bf Q}\cdot \langle n,{\bf Q}|i\nabla_{\bf Q} |n,{\bf Q}\rangle.
\label{chi}
\ee
Physically, as ${\bf Q}$ changes, the electronic ket $|\psi\rangle$ stays ``parallel'' to its
initial direction (in the sense of Levi-Civita), while the local reference ket $|n,{\bf Q}\rangle$ 
rotates relative to it.  The Berry phase is the phase angle between them.

The form of Eq. \ref{chi} suggests that it is fruitful to view the integrand as a (Berry) vector potential
${\bf A(Q}) = \langle n,{\bf Q}|i\nabla_{\bf Q} |n,{\bf Q}\rangle$.  We may then regard $\chi(C)$ 
as an Aharanov-Bohm (AB) phase caused by an effective magnetic 
field ${\bf B(Q}) = \nabla_{\bf Q} \times {\bf A(Q})$ that lives in parameter 
space (${\bf B(Q)}$ is called the Berry curvature).

\section{Anomalous velocity}\label{anomalous}
For an electron in a periodic potential, the eigenstates are the Bloch states 
\be
\psi_{n\bf k}({\bf r}) = \frac{1}{\sqrt N} \mathrm{e}^{i\bf k\cdot r}u_{n{\bf k}}({\bf r}) = \langle {\bf r}|n{\bf k}\rangle.
\label{bloch}
\ee
If we ignore the spin, the band index $n$ and wave-vector ${\bf k}$ are sufficient 
to fully characterize its state.  

Let us perturb the electron by adding a static potential $V({\bf r})$.  In Bloch representation, the Hamiltonian is
\be
H = V({\bf R}) + \epsilon_n({\bf k}),  
\label{H1}
\ee
where ${\bf R}= i\nabla_{\bf k}$ is the Wannier coordinate indexing the lattice sites and $\epsilon_n({\bf k})$
is the unperturbed band energy.  The potential $V$ causes 
the wave vector ${\bf k}$ to drift along a path $C$ in reciprocal space at the rate
\be
\hbar{\bf \dot{k}} = -i[{\bf k}, H] =  -\frac{\partial V} {\partial {\bf R}}.
\label{kdot}
\ee
In principle, transitions to other bands $n'\neq n$ will occur with finite amplitude.  However, for weak $V$, it is 
customary to assume that the electron always remains in the same band $n$.  This assumption 
is a \emph{constraint} analogous to those in the preceding examples.  
Hence we should expect the electron's ket vector $|\psi\rangle$ to acquire a Berry phase, i.e. along $C$,
$\langle {\bf r}|\psi\rangle = u_{n{\bf k}}({\bf r})\mathrm{e}^{-i\chi({\bf k})}$ where
\be
\chi({\bf k}) = -\int^{\bf k}_C d{\bf k}'\cdot {\bf X}({\bf k}'), 
\label{chi2}
\ee
with the Berry vector potential ${\bf X(k)}$ given by
\be
{\bf X}({\bf k}) = \int_{cell} d^3r\; u^*_{n{\bf k}}({\bf r})i\nabla_{\bf k}u_{n\bf k}({\bf r})
\label{X}
\ee
(${\bf X}$ has dimensions of length).
With the substitution ${\bf k}\rightarrow {\bf Q}$,  Fig. \ref{hilbert}b equally well depicts
the parallel transport of $|\psi\rangle$ as ${\bf k}(t)$ changes in reciprocal space.

As in Eq. \ref{chi}, we may regard $\chi$ as the AB phase caused by a magnetic field 
\be
{\bf\Omega(k)} = \nabla_{\bf k} \times {\bf X(k)}
\label{Omega}
\ee
that lives in $\bf k$ space.

The Berry curvature ${\bf \Omega}$ alters the response of the electron to $V$
in an essential way.  To see this, we perform a gauge transformation to remove the 
AB phase at the cost of adding the vector potential $\bf X$ to $i{\bf \nabla_{k}}$ in 
the argument of $V$ in Eq. \ref{H1}.  The transformed Hamiltonian is then
\be
H' = V(i{\bf \nabla_{k}} + {\bf X}) + \epsilon_n({\bf k}).
\label{H2}
\ee
There are 2 ways to view $H'$.  The role of ${\bf X(k)}$ as a vector gauge potential is now manifest.  
Moreover, in the argument of $V$, the position operator is now given by
\be
{\bf x} = {\bf R + X(k)}
\label{xop}
\ee
instead of just ${\bf R}$.  We may view ${\bf X(k)}$ as an intracell coordinate that locates the 
electron within the unit cell.  Equation \ref{xop} implies that $\bf x$ fails to commute with itself.  Instead
we have~\cite{Adams}
\be
[x_i,\; x_j] = i\epsilon^{ijk}\Omega_k,
\label{xx}
\ee
with $\epsilon^{ijk}$ the totally antisymmetric tensor.

With the Hamiltonian $H'$, $\hbar{\bf\dot{k}}$ is unchanged 
from Eq. \ref{kdot}, but the group velocity is (using Eq. \ref{xx})
\be
\hbar{\bf v} = -i[{\bf x}, H'] = \nabla_{\bf k} \epsilon_n({\bf k}) + 
\left(\frac{\partial V}{\partial {\bf x}} \right) \times {\bf \Omega(k)}.
\label{v0}
\ee
The extra term involving ${\bf \Omega}$ is called the (Luttinger) anomalous velocity.

If $V = -e{\bf E\cdot x}$ ($e$ is the elemental charge and $\bf E$ an electric field), the semiclassical equations of motion 
become (we now include a weak, physical magnetic field ${\bf B}$ for completeness)
\begin{eqnarray}
\hbar{\bf\dot{k}} &=& e{\bf E} + e{\bf v\times B}\label{kdot1}\\
\hbar{\bf v} &=& \nabla \epsilon_n - e{\bf E\times \Omega}.\label{v1}
\end{eqnarray}
The Luttinger term $e{\bf E\times \Omega}$ brings a rather pleasing symmetry between ${\bf k}$ and
${\bf R}$ space.  
 
\section{Anomalous Hall current}\label{AHE}
In the Boltzmann-equation approach, the charge current density in the absence of $\bf B$ is
\be
{\bf J} = e\sum_{{\bf k},s} {\bf v_k} [f^0_{\bf k} + g_{\bf k}],
\label{J}
\ee
where $f^0_{\bf k}$ is the equilibrium distribution function
and $g_{\bf k}= e{\bf v_k\tau\cdot E} (-\frac{\partial f^0}{\partial \epsilon})$
the correction caused by $\bf E ||\hat{x}$ ($\tau$ is the 
transport relaxation time).  Normally, the term in $f^0_{\bf k}$ sums to zero by symmetry, as it must, leaving 
the term in $g_{\bf k}$ to give the usual conductivity $\sigma$ (the Hall conductivity $\sigma_{xy}$ = 0).  Now, with 
$\bf v_k$ given by Eq. \ref{v1}, the term in $f^0_{\bf k}$ yields a sizeable current transverse to $\bf E$, viz.
\be
{\bf J}_H = \frac{e^2}{\hbar} {\bf E\times \sum_{{\bf k},s}\Omega(k)} f^0_{\bf k}
\label{Jy}
\ee
Because it derives from $f^0_{\bf k}$, ${\bf J}_H$ has the remarkable feature that it is
\emph{independent} of $\tau$, as found by KL~\cite{Karplus}.  We may write the AHE
conductivity as 
\be
\sigma'_{xy} = n\frac{e^2}{\hbar} \langle\Omega\rangle,
\label{sxy}
\ee
where $\langle \Omega\rangle \equiv n^{-1} \sum_{\bf k} \Omega_z({\bf k}) f^0_{\bf k}$ is the weighted
average of the Berry curvature.

It turns out that, if time-reversal invariance holds, as in a non-magnetic conductor, $\bf \Omega(k)$ vanishes.  
In a ferromagnet, the uniform magnetization $\bf M$ breaks time-reversal symmetry 
for the spins.  This symmetry-breaking is communicated to the charge currents via 
spin-orbit coupling.  In the simple case of a ferromagnetic
semiconductor, Nozieres and Lewiner (NL)~\cite{NL} have calculated
${\bf X(k)} = -\lambda {\bf k\times S}$,
where $\lambda$ is the spin-orbit coupling parameter and $\bf S\sim M$.  Equation \ref{Jy} then gives
for the NL model an AHE current 
${\bf J}_H = 2ne^2\lambda {\bf E\times S}$
that is linear in the carrier density $n$ and $\bf M$, but independent of $\tau$.  

Generally, the Hall resistivity $\rho'_{xy} = \sigma_{xy}\rho^2$ is the quantity that is measured.  
Since $\rho\sim(n\tau)^{-1}$, Eq. \ref{sxy} immediately implies that~\cite{Lee1}
\be
\rho'_{xy} = An\rho^2
\label{rhoxy}
\ee
with $\rho$ the resistivity and $A = e^2\langle\Omega\rangle/\hbar$.  
In an experiment in which both $n$ and $\tau$ can be varied, the KL 
theory predicts that $\rho'_{xy}/n$ scales like $\rho^{\alpha}$ with $\alpha$ = 2.  By contrast, skew-scattering theories
predict $\sigma_{xy}\sim n\tau$ so that $\rho'_{xy}$ is linear in $\rho$.

\section{Spinel ferromagnet}\label{spinel}
These predictions have to be tested in the impurity-scattering regime because,
if $\tau$ is dominated by phonon/magnon scattering, both theories predict $\alpha$ = 2. 
Moreover, it would be desirable to change $n$ greatly (in addition to 
$\tau$) without destroying the magnetization.  These conditions are met in the spinel ferromagnet 
$\rm CuCr_2Se_{4-x}Br_x$ which is a metal with a Curie temperature $T_C\sim$ 400 K (for $x$ = 0)~\cite{Lee1}. 
Interaction between the local moments on adjacent Cr ions is ferromagnetic because the Cr--Se--Cr bond 
is 90$^\mathrm{o}$ (this suppresses the superexchange term ${\cal J}_S$ to leave only direct exchange).
The holes which reside on the Se bands contribute negligibly to the exchange.
Hence tuning the hole density $n$ (by changing the Br content $x$) hardly affects the magnetization.
As $x$ increases from 0 to 1, $n$ decreases from $7\times 10^{21}$ to $\sim 10^{20}$ 
cm$^{-3}$, while $\tau$ decreases by a factor of 70 to give a change in resistivity $\rho$ exceeding
3 decades (measured at 5 K)~\cite{Lee1}.  However, $M$ at 5 K remains nearly unchanged ($T_C$ decreases to 250 K). 
Typical profiles of $\rho_{xy}$ vs. $H$ are shown in Fig. \ref{Hall}.

\bfig[h]			
\incl[width=5.7cm]{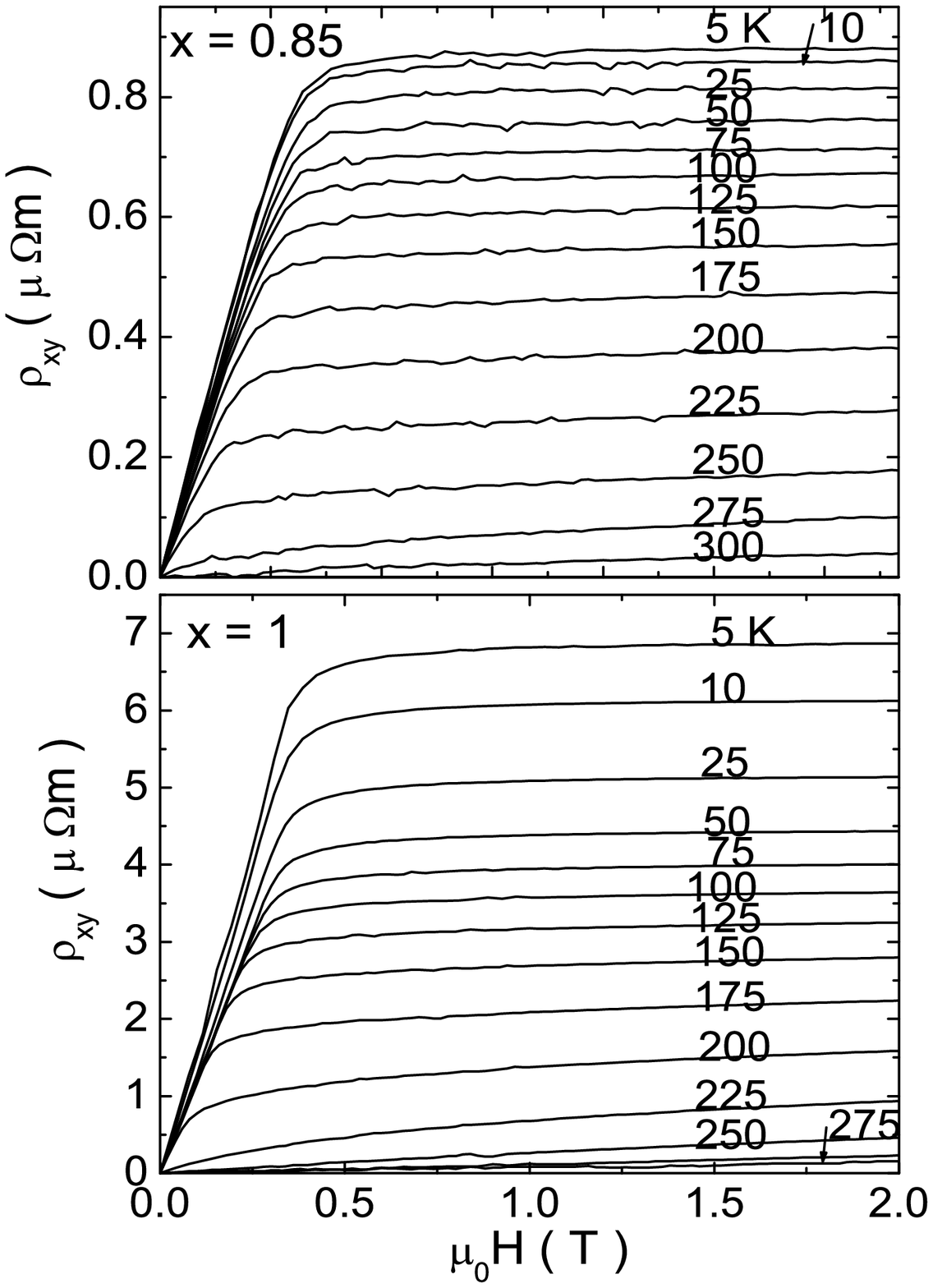}
\caption{\label{Hall} Traces of the observed $\rho_{xy}$ vs. $H$ in $\rm CuCr_2Se_{4-x}Br_x$ 
for $x$ = 0.85 (top panel) and 1.0 (bottom)~\cite{Lee1}.  At each $T$, $\rho_{xy}$ varies like $M$ vs. $H$ ($R_0$ is negligible).  
In the $x$= 1.0 sample, $\rho'_{xy}$ at 5 K is possibly the largest AHE resistivity observed to date in a ferromagnet.
}
\efig 

\bfig[h]			
\incl[width=7cm]{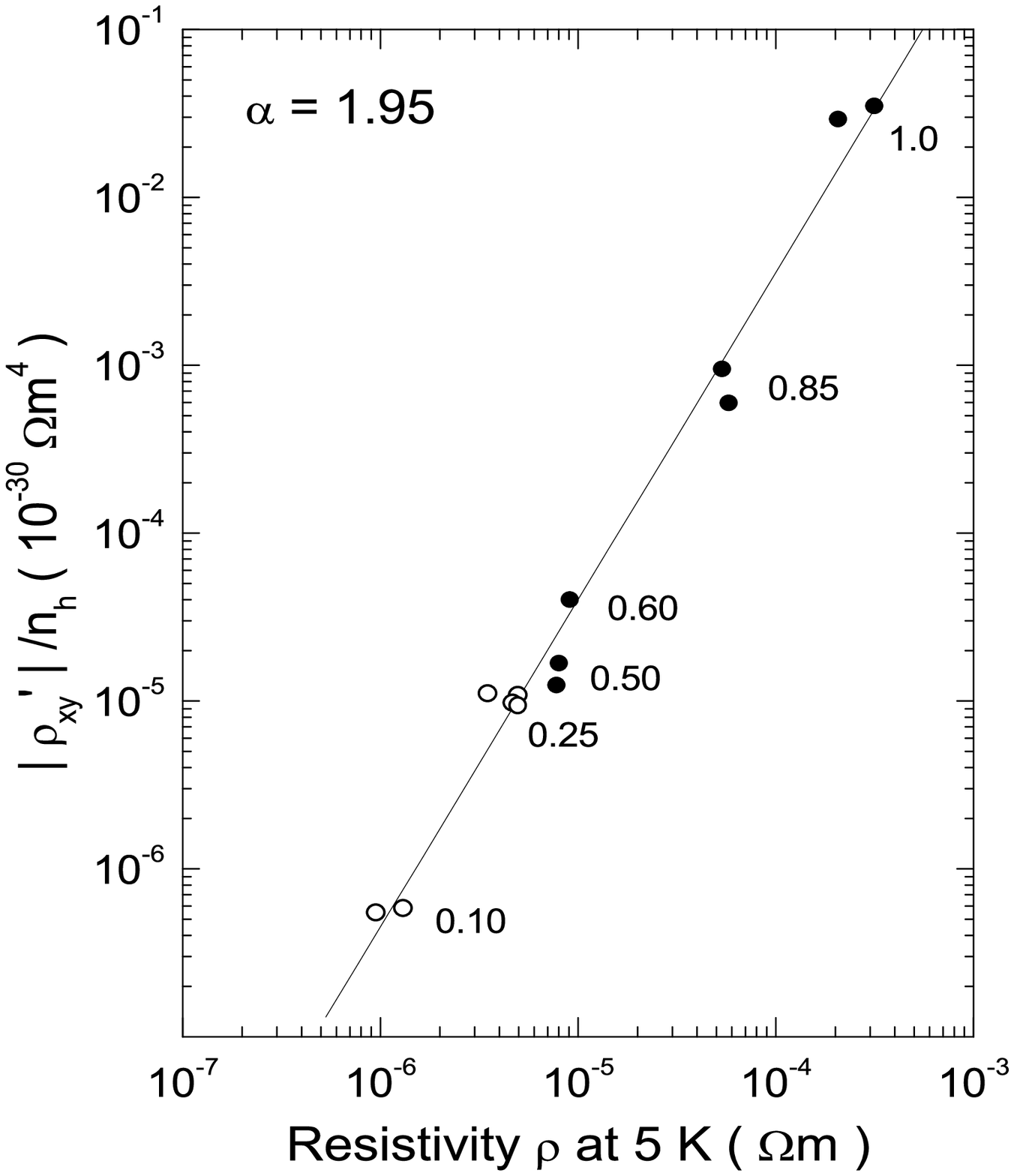}
\caption{\label{loglog} Plot of the saturation value of $\rho'_{xy}/n$ versus $\rho$ (both measured at 5 K) in log-log scale
for 12 crystals of $\rm CuCr_2Se_{4-x}Br_x$~\cite{Lee1}.  Open (solid) circles indicate negative (positive) sign for $\rho'_{xy}$
at 5 K ($\rho'_{xy}$ for the $x$ = 0 sample is not resolved).  The solid line is a fit 
to $|\rho'_{xy}|/n = A\rho^{\alpha}$ with $\alpha$ = 1.95 and $A = 2.24 \times 10^{-25}$ (SI units).
}
\efig 

In Fig. \ref{loglog}, $|\rho'_{xy}|/n$ measured in 12 crystals at 5 K is plotted against $\rho$.
As $\rho$ varies over 3 decades, $\rho'_{xy}/n$ increases
by 6 decades, following Eq. \ref{rhoxy} with $\alpha$ = 1.95.  This implies that $\sigma'_{xy}$ is 
linear in $n$, but independent of $\tau$.  However, the sharp sign change
near $x$ = 0.4 is not understood at present.  [To compare with the skew-scattering prediction, we have 
also plotted $\rho'_{xy}$ against $\rho$.  The data yield an exponent of 1.36
instead of 1.0.  The disagreement with skew scattering lies well outside the uncertainties of the data.]

The experiment provides an estimate of the average Berry curvature $\langle\Omega\rangle$.  Using Eq. \ref{sxy},
we find that the parameter $A= 2.24\times 10^{-25}$ (SI units) gives 
$\langle\Omega\rangle^{\frac12}$ = 0.30 \AA.  

\section{Anomalous Nernst Effect}\label{ANE}
The AHE corresponds to a charge current that flows transverse to $\bf E$.  In addition to the charge, the electron current
carries heat.  This may be observed by the anomalous Nernst effect~\cite{Lee2}.  In the Nernst experiment, an applied  
temperature gradient $-\nabla T || \bf\hat{x}$ produces a transverse charge current which is antisymmetric in
$\bf H||\hat{z}$, viz. $J_y = \alpha_{yx}(-\nabla T)$, where $\alpha_{ij}$ is the Peltier conductivity tensor.
With the Luttinger velocity, we have
\be
\alpha_{xy} = e\sum_{{\bf k},s} \frac{\epsilon_{\bf k}-\mu}{T} 
\left(-\frac{\partial f^0}{\partial \epsilon} \right) v^0_xk_x\Omega_z({\bf k}),
\label{alpha}
\ee
where $k_B$ is Boltzmann's constant, $\mu$ the chemical potential, and $\hbar{\bf v}^0 =\nabla \epsilon({\bf k})$.
In deriving this, we have made the substitution $\hbar{\bf k}/\tau\rightarrow e{\bf E}$ in Eq. \ref{v1}.  Defining
the quantity $\langle\Omega\rangle_{\epsilon}$ as the angular average of $\bf\Omega(k)$ over the Fermi Surface (FS),
we may simplify Eq. \ref{alpha} to
\be
\alpha_{xy} \rightarrow \frac{\pi^2}{3} \frac{ek_B^2T}{\hbar}\;\frac23  
\left(\frac{\partial \langle\Omega\rangle_{\epsilon}\epsilon{\cal N}}{\partial \epsilon} \right)_{\mu}.
\label{alpha1}
\ee
With the density of states ${\cal N}\sim\sqrt{\epsilon}$, the last factor reduces 
to $\Omega{\cal N}_F$ if the $\epsilon$-dependence of $\Omega$ is negligible.  

In a recent experiment~\cite{Lee2} in which the thermopower, Hall angle and $\rho$
were combined with the Nernst effect to extract $\alpha_{xy}$ for 5 of the crystals of $\rm CuCr_2Se_{4-x}Br_x$ 
used in the AHE experiment, it was found that at low $T$
\be
\alpha_{xy} = {\cal A}\frac{ek_B^2T}{\hbar}{\cal N}^0_F
\label{fit}
\ee
with ${\cal N}^0_F$ the free-electron density of states, and ${\cal A}\sim$34 \AA$^2$.  The agreement with Eq. \ref{alpha1}
is encouraging although the numerical coefficient seems too large to be simply identified with $\Omega$.  
We note, however, that the free-electron ${\cal N}^0$ used in Eq. \ref{fit} implies that ${\cal A}$ 
is enhanced by the true effective mass.  The 2 experiments probe different ways of
averaging $\bf \Omega(k)$ over $\bf k$.

We acknowledge valuable discussions with N. Nagaosa, F. D. M. Haldane, R. J. Cava and G. Sawatzky, and 
support from the U. S. National Science Foundation through a MRSEC grant (DMR 0213706).

$^\dagger$\emph{Present address of WLL: Johns Hopkins Univ, Dept. Phys. and Astron., Baltimore, MD 21218, USA}
\end{document}